Generalized Paraxial Ray Tracing Derived from Geodesic Deviation

David R Bergman[a]

**Abstract**

Paraxial ray tracing procedures have become widely accepted techniques for acoustic models in seismology and underwater acoustics. To date a generic form of these procedures including fluid motion and time dependence has not appeared in the literature. A detailed investigation of the characteristic curves of the equations of hydrodynamics allows for an immediate generalization of the procedure to be extracted from the equation for geodesic deviation. The general paraxial ray trace equations serve as an ideal supplement to ordinary ray tracing in predicting the deformation of acoustic beams in random environments. The general procedure is derived in terms of affine parameterization and in a coordinate time parameterization ideal for application to physical acoustic ray propagation. The formalism is applied to layered media, where the deviation equation reduces to a second order differential equation for a single field with a general solution in terms of a depth integral along the ray path. Some features are illustrated through special cases which lead to exact solutions in terms of either ordinary or special functions.

[a] Saint Peter's College, 2641 Kennedy Boulevard, Jersey City, New Jersey 07306
    dbergman@spc.edu

# I. INTRODUCTION

Paraxial ray tracing has become widely used in Seismology where motion of the medium can be ignored. In paraxial ray trace procedures ray paths within a ray bundle or tube are determined by solving a system of second order linear differential equations along one ray path rather than repeatedly solving the ray equation with different initial conditions [1]. A complete set of paraxial ray equations for three dimensional acoustic ray tracing in an arbitrary environment are presented. From the results of this procedure the neighboring rays close to a specific solution of the ray equations may be mapped thus allowing one to model the deformation of a ray bundle as it propagates through a random environment. The equation for geodesic deviation, commonly used in differential geometry to study the existence of conjugate points along geodesics of a differential manifold, is used to derive the generic paraxial ray trace procedure [2], [3], [4].

It is worth noting that while the identification of acoustic rays with null geodesics was first illustrated by R. White in 1973 some ten years later the connection was discovered again in the general relativity community and has become known as acoustic GR (or acoustic analog models of black holes) [5], [6]. The approaches taken in each case identify the acoustic rays as paths of zero length in four dimensional space time called null geodesics. A significant amount of research has been done using these equations to model the sound in the presence of moving fluids [7], [8], [9].

New techniques for solving the parabolic wave equation in ducted environments based on operator expansions and path integrals techniques originally developed and used in quantum field theory have been employed by practitioners in the acoustics community with great success [10], [11]. These techniques have led to the development of new approaches to numerical modeling of sound in ocean environments. This work follows this modern tradition in that the use of geodesic deviation in general relativity serves as an inspiration.

The presence of differential geometry in modern physics for the past 75 years has led to the development and use of many sophisticated techniques that are employed in the development of the paraxial approximation and the analysis of layered media presented here. While much of the formalism can be extracted from texts on general relativity and mathematical physics the geometric relationship between the four dimensional deviation and the deviation between equal time rays used later in this work, while simple and elegant, is not trivial and a derivation of the system is presented in this article for completeness. The reader is referred to the classic text book literature in these fields for more information [2], [12-15].

This article is divided into four sections, section 1 being the introduction. Section 2 introduces the mathematical machinery necessary for constructing a paraxial ray trace algorithm in a random fluid medium and the geometrical significance of the quantities involved. This system consist of the ray equation, the equations for determining a ray centered basis parallel propagated along the ray and the equation of geodesic deviation expressed in the ray centered basis. The paraxial system is initially derived using an affine parameter, an obvious choice from the geometric point of view, and later recast in a more tractable form using coordinate time as a three dimensional ray parameter. In section 3 the procedure is applied to layered media with fluid motion where the technique of identifying isometries of the metric is used to determine the ray tangent at all point along the ray. These results are then fed into the deviation equation with attention being focused on effective two dimensional ray systems where it is demonstrated that the equations of parallel transport need not be solved. Hence a ray centered basis may be chosen by inspection resulting in a scalar equation for a single component of the deviation vector. The effects of environmental parameters on the focusing properties of the medium are discussed in general and some exact solutions presented. Although a complete recipe for the construction of a numerical procedure is presented in its entirety the applications presented in this article are geared mainly towards situations where either an exact solution may be derived or general properties inferred without explicit numerical or analytical results. Since examples of problems with exact solutions are rare their presentation serves both a research and pedagogical value towards acoustic modeling and understanding the techniques involved. In section four the conclusions are summarized and discussed.

The standard conventions of tensor analysis on a differentiable manifold are used throughout; the reader is referred to the literature for more detailed information [2]. Covariant tensors are denoted $A_\mu$ contravariant tensors $A^\mu$. Indices are raised and lowered by the metric tensor $A_\nu = g_{\nu\mu} A^\mu$, and the Einstein summation convention is used throughout (whenever the same index appears twice, once in the covariant position and once in the contravariant position, sum over all index values is implied). The index convention for tensor components refers to a local Cartesian coordinate patch with the coordinates of a point in the space (or space - time) given by $x^\mu$. The author uses a different signature for the metric tensor than White, following the trend commonly found in general relativity and much of the standard jargon used throughout is common in general relativity. The metric tensor $g_{\mu\nu}$ provides a generalization of the concept of a "dot" product, which calculates the magnitude of a vector and the angle between two vectors in a tangent space at each point of the manifold. The index notation used throughout serves a dual purpose in labeling tensor components with respect to the manifold with inner product $g_{\mu\nu}$ and components of vector quantities in ordinary Euclidian space, such as the fluid velocity $\upsilon_j$ (Greek indices run from 0…3 and Latin from 1…3, with "0" reserved for time). In the first case indices may be lowered and raised by the metric tensor and its inverse respectively, while in the latter case there is no difference between covariant and contravariant notation, $\upsilon_j = \upsilon^j$. Although the formal derivation begins by considering rays as four dimensional space–time paths of a differentiable manifold all quantities are reduced to ordinary three dimensional vectors in Euclidian space to illustrate the connection between this approach and that appearing in Cerveny. Describing the relation between four dimensional space-time vectors and ordinary Euclidian vectors within this formalism requires a shift from one language to another. Since this shift in notation will occur often throughout this paper the paradigm used in a specific equation or derivation will always be noted in the text by the following shorthand; a differentiable manifold endowed with an inner product on each tangent space is referred to simply as *M*, ordinary Euclidian space with the standard "dot product" is called $E^3$.

## II. DERIVATION OF THE PARAXIAL RAY TRACE PROCEDURE

The relationship between the bicharacteristics of a partial differential equation and the geodesic flow of a Riemannian manifold is well documented in the literature. The presentation of the following work follows closely that found in two references, Courant & Hilbert, Methods of Mathematical Physics Volume II and Fritz John, Partial Differential Equations, New York University, [17] and [18] respectively. The equations describing the propagation of acoustic wave fronts in the weak field limit were derived from the results given in Courant & Hilbert by R. J. Thompson [8]. Thompson's presentation explicitly separates the fluid velocity into a "wind" vector and an acoustic perturbation although the pressure and density fields are not expanded. In this treatment the characteristics and bicharacteristics for the complete set of equations of hydrodynamics is discussed. Thus the approach presented here is more general than those of past authors. In particular it is shown that the characteristics of the full field equations are identical to the null geodesic of a Lorentzian manifold similar to that studied in general relativity.

The equations of hydrodynamics consists of the continuity equation and Euler's equation,

$$\partial_t \rho + \vec{v} \cdot \vec{\nabla} \rho + \rho \vec{\nabla} \cdot \vec{v} = 0 \tag{1}$$

$$\partial_t \vec{v} + \vec{v} \cdot \vec{\nabla} \vec{v} + \frac{1}{\rho} \vec{\nabla} p = 0, \tag{2}$$

where $p$ is the fluid pressure, $\vec{v}$ the fluid velocity and $\rho$ the fluid density. In addition it is assumed that the fluid obeys an equation of state $\rho = \rho(p)$ with $\rho' = c^{-2}$ defining the local sound speed. Using the equation of state in Euler's equation, equation 2, leads to the following set of four partial differential equation for the four fields $\rho$, $\vec{v}$

$$\partial_t p + \vec{v} \cdot \vec{\nabla} p + \rho c^2 \vec{\nabla} \cdot \vec{v} = 0 \tag{5}$$

$$\partial_t \vec{v} + \vec{v} \cdot \vec{\nabla} \vec{v} + \frac{1}{\rho} \vec{\nabla} p + \vec{\nabla} \Phi = 0, \tag{6}$$

where $\Phi$ is a potential due to external forces such as gravity acting on the fluid. Equations 5 and 6 will be used as the starting point for the derivation of the characteristics and bicharacteristics of this system. Equations 5 and 6 are a system of first order quasi-linear partial differential equations. The characteristic matrix for the set of equations is

$$\begin{bmatrix} D\varphi & \rho c^2 (\vec{\nabla}\varphi)^T \\ \rho^{-1}\vec{\nabla}\varphi & 1_{3\times 3} D\varphi \end{bmatrix}, \tag{7}$$

where $D \equiv \partial_t + \vec{v} \cdot \vec{\nabla}$ and $(\vec{\nabla})^T \equiv (\partial_x \quad \partial_y \quad \partial_z)$ is a short hand notation for a row vector with entries equal to the components of the gradient.

The characteristic conditions is satisfied be setting the determinant of the matrix equal to zero leading to,

$$Q = (D\varphi)^2 \left((D\varphi)^2 - c^2 \vec{\nabla}\varphi \cdot \vec{\nabla}\varphi\right) = 0. \tag{8}$$

Each term in equation 8 defines a partial Hamintonian for the system of equations and may be treated separately. Defining $H_1 = (D\varphi)^2$ and $H_2 = \left((D\varphi)^2 - c^2 \vec{\nabla}\varphi \cdot \vec{\nabla}\varphi\right)$ one set of characteristics is determined by

$$H_2 = g^{\mu\nu} \partial_\mu \varphi \partial_\nu \varphi = 0, \tag{9}$$

where the contravariant metric tensor $g^{\mu\nu}$ has been introduced with $g^{00} = -c^{-2}$, $g^{0i} = -\mu_i$, $g^{ij} = \delta_{ij} - \mu_i \mu_j$, with $\bar{\mu} \equiv \bar{v}/c$ defining the local Mock number, or in matrix form

$$g^{\mu\nu} \equiv \frac{1}{c^2} \begin{pmatrix} -1 & -v_i \\ -v_j & c^2 \delta_{ji} - v_j v_i \end{pmatrix}. \tag{10}$$

The corresponding covariant metric tensor given by

$$g_{\mu\nu} = \begin{pmatrix} -c^2 + v^2 & -v_i \\ -v_j & \delta_{ji} \end{pmatrix} \tag{11}$$

is the inverse of equation 10, $g_{\mu\alpha} g^{\alpha\nu} = \delta_\mu^\nu$. A factor of $\rho c^2$ has been removed from equation… for simplicity but otherwise having no effect on the geometry of the curves defined by equation…. This reflects the well known fact that ray theory, more generally the bi-characteristic geometry, is immune to the acoustical impedance of the fluid. The bi-characteristics, which are identified with the curves in space–time along which energy is propagated, or with the rays of the system, are given by a set of trajectories in space time. Defining a parameter $\lambda$ along these curves, their tangent vectors are related to $\partial_\mu \varphi$ by the equation

$$\frac{dx^\mu}{d\lambda} = g^{\mu\nu} \partial_\nu \varphi. \tag{12}$$

The characteristic and the bi-characteristic curves obey a set of ordinary first order differential equations determined by the characteristic condition $H_2 = 0$. Replacing $p_\mu \equiv \partial_\mu \varphi$ in $H_2$ the equations for the system are

$$\frac{dx^\mu}{d\lambda} = \frac{1}{2}\frac{\partial H_2}{\partial p_\mu} = g^{\mu\nu} p_\nu, \tag{13}$$

$$\frac{dp_\mu}{d\lambda} = -\frac{1}{2}\frac{\partial H_2}{\partial x^\mu} = -\frac{1}{2} p_\alpha p_\beta \partial_\mu g^{\alpha\beta}. \tag{14}$$

Equation 13 simply relates the velocity of the bicharacteristic to the wave-front gradient which is identified here with the conjugate momentum to the variable $x^\mu$ and may be inverted using equation 11 to yield $p_\mu = g_{\mu\nu}\dot{x}^\nu$, where $\dot{x}^\nu \equiv dx^\nu/d\lambda$. In equation 14 the derivative of the contravariant metric can be replaced by the derivatives of the covariant metric by differentiating $g_{\mu\alpha} g^{\alpha\nu} = \delta_\mu^\nu$, thus combining equations 13 and 14 leads to the following equation for the bicharacteristics,

$$\frac{d^2 x^\mu}{d\lambda^2} + \frac{1}{2} g^{\mu\nu} \{2\partial_\alpha g_{\nu\beta} - \partial_\nu g_{\alpha\beta}\} \frac{dx^\alpha}{d\lambda}\frac{dx^\beta}{d\lambda} = 0, \tag{15}$$

which is precisely the equation for the geodesics of a differentiable manifold $M$ with metric defined by equation 11. From equation 13 and the characteristic condition it follows that the bicharacteristics of equation 5 and 6 are equivalent to the null geodesics defined by the metric tensor and its inverse, equations 11 and 10. Equation 15 is equivalent to the Euler – Lagrange equations derived from an action of the form, $\int L d\lambda$ with $L \equiv g_{\mu\nu}\dot{x}^\mu \dot{x}^\nu = p_\mu \dot{x}^\nu$.

In practice one distinguishes between bulk fluid motion and acoustic disturbances propagating within the bulk describing each separately. To describe acoustics in the presence of background fluid motion all fields in the problem are written in the form $\vec{v} = \vec{w} + \vec{v}$, $p = p_0 + p_1$ and $\rho = \rho_0 + \rho_1$ where $\vec{w}$, $p_0$ and $\rho_0$ are the velocity, pressure and density of the fluid medium and $\vec{v}$, $p_1$ and $\rho_1$ are contributions due to acoustic phenomenon. In the following treatment of the problem it is assumed that the perturbations are weak, leading to a linear theory for the propagation of sound. No restriction is placed on the state equation for the background fluid motion while the acoustic fluctuations are assumed to obey the adiabatic condition, $\rho_1 = c^{-2} p_1$ where $c^{-2} \equiv \partial\rho/\partial p|_0$ defines the local sound speed. The background fields are assumed to obey the equations of hydrodynamics; with body forces such as gravity are lumped together in the background equations, independently of the acoustic field. Applying these assumptions to equations 1 and 2

$$\{D^* \rho_0 + \rho_0 \vec{\nabla}\cdot\vec{w}\} + \{c^{-2} D p_1 + \rho\vec{\nabla}\cdot\vec{v}\} + \{\rho_1 \vec{\nabla}\cdot\vec{w} + \vec{v}\cdot\vec{\nabla}\rho_0 + p_1 D c^{-2}\} = 0 \tag{16}$$

$$\{\rho_0 D^* \vec{w} + \vec{\nabla} p_0\} + \{\rho D\vec{v} + \vec{\nabla} p_1\} + \{\rho_1 D^* \vec{w} + \rho\vec{v}\cdot\vec{\nabla}\vec{w}\} = 0 \tag{17}$$

where $D \equiv \partial_t + \vec{v} \cdot \vec{\nabla}$ as before, $D^* \equiv \partial_t + \vec{w} \cdot \vec{\nabla}$ following Thompson's notation and the adiabatic assumption has been used to eliminate derivatives of $\rho_1$. The first terms in equations 16 and 17 vanish by the assumption placed on the fluid motion leading to the equations for the acoustic fields.

$$\{D^* p_1 + \rho_0 \vec{\nabla} \cdot \vec{v}\} + \{p_1 \vec{\nabla} \cdot \vec{w} + c^2 \vec{v} \cdot \vec{\nabla} \rho_0 + p_1 c^2 D^* c^{-2}\} = 0 \tag{18}$$

$$\{\rho_0 D^* \vec{v} + \vec{\nabla} p_1\} + \{c^{-2} p_1 D^* \vec{w} + \rho_0 \vec{v} \cdot \vec{\nabla} \vec{w}\} = 0 \tag{19}$$

where only linear terms in the acoustic field appear. It is not assumed that the background quantities or there variations are small.

The characteristic condition for equations 18 and 19 is determined only by the highest order differential operator appearing in the equations. Let the four component vector $u^{[i]}$ denote the field variables with $u^{[i]} \equiv v_i$ for $i = 2, 3, 4$ and $u^{[1]} \equiv p_1$, then equations 18 and 19 may be written in matrix form

$$L_{(1)}[u] + M \cdot u = 0 \tag{20}$$

where $L_{(1)}[\cdots]$ is a first order differential operator acting on $u$ while $M$ is an ordinary matrix acting on $u$. In both cases these matrices depend on the background fields and the acoustic fields. From equations 18 and 19 the explicit form of $L_{(1)}[\cdots]$ and $M$ are

$$L_{(1)} = \begin{bmatrix} D^* & \rho_0 c^2 (\vec{\nabla})^T \\ \vec{\nabla} & 1_{3 \times 3} \rho_0 D^* \end{bmatrix} \tag{21}$$

$$M = \begin{bmatrix} c^{-2} \vec{\nabla} \cdot \vec{w} + D^* c^{-2} & (\vec{\nabla} \rho_0)^T \\ c^{-2} D^* \vec{w} & \rho_0 \vec{\nabla} \otimes \vec{w} \end{bmatrix} \tag{22}$$

where $\vec{\nabla} \otimes \vec{w}$ represents a $3 \times 3$ matrix with components $\partial_i w_j$. One can see by inspection that the first terms in equations 18 and 19 resemble equations 5 and 6 leading to the conclusion that the characteristic condition for acoustic perturbations in the presence of background fluid motion is given by equation 9, where the terms in the metric contain the background $\vec{w}$ instead of $\vec{v}$. The fluid density factors out and may be dropped without loss of generality.

One may derive second order wave equations for the pressure and velocity field directly from equations 5 and 6. Applying $D$ to equation 5 after dividing through by $\rho c^2$, taking the divergence of equation 6 and taking the difference of the resulting equations leads to

$$D \frac{1}{\rho c^2} Dp - \vec{\nabla} \cdot \frac{1}{\rho} \vec{\nabla} p + \Delta_1 = 0, \tag{20}$$

while taking the gradient of equation 5 and applying $D$ to equation 6 leads to

$$D\rho D\bar{v} - \bar{\nabla}\rho c^2 \bar{\nabla}\cdot\bar{v} + \bar{\Delta}_2 = 0, \tag{21}$$

$\Delta_1 \equiv -\partial_k v_j \partial_j v_k$ and $\bar{\Delta}_2 \equiv -\partial_k p \bar{\nabla} v_k$.

The steps leading to equations 20 and 21 are similar to those taken in deriving a second order wave equation from Maxwell's equations for the electromagnetic field tensor and the approach taken by White [7]. Applying the same procedure to equation 18 and 19 leads to a similar result for the linear acoustic field in the presence of background motion.

$$\left\{ D^* \frac{1}{\rho_0 c^2} D^* p_1 - \bar{\nabla}\cdot\frac{1}{\rho_0}\bar{\nabla}p_1 \right\} + \tilde{\Delta}_1 = 0, \tag{21}$$

$$\left\{ D^* \rho_0 D^* \bar{v} - \bar{\nabla}\rho_0 c^2 \bar{\nabla}\cdot\bar{v} \right\} + \tilde{\bar{\Delta}}_2 = 0, \tag{22}$$

$$\tilde{\Delta}_1 \equiv D^*\left(\frac{p_1}{\rho_0 c^2}\bar{\nabla}\cdot\bar{w}\right) - \bar{\nabla}\cdot\left(\frac{p_1}{\rho_0 c^2}D^*\bar{w}\right) + D^*\left(\bar{v}\cdot\bar{\nabla}\ln\rho_0\right) + D^*\left(\frac{p_1}{\rho_0}D^*c^{-2}\right)$$
$$-\partial_k\left(\bar{v}\cdot\bar{\nabla}w_k\right) - \partial_k v_j \partial_j w_k$$

$$\tilde{\bar{\Delta}}_2 \equiv D^*\left(\rho_0\bar{v}\cdot\bar{\nabla}\bar{w}\right) + D^*\left(\rho_1 D^*\bar{w}\right) - \bar{\nabla}\left(p_1\bar{\nabla}\cdot\bar{w}\right) - \bar{\nabla}\left(c^2\bar{v}\cdot\bar{\nabla}\rho_0\right) - \partial_k p_1 \bar{\nabla}w_k.$$

The equation for the Eikonal in the high frequency limit of linear acoustics, like that for the bi-characteristics, comes from the highest order derivatives. The terms $\tilde{\Delta}_1$ and $\tilde{\bar{\Delta}}_2$ contain only the field variables and first order derivatives. The high frequency plane wave approximation is defined by taking $\alpha \to 0$, $p_1 = \pi_\alpha e^{i\varphi/\alpha}$ and $\bar{v} = \bar{\sigma}_\alpha e^{i\varphi/\alpha}$ with $\pi_\alpha \equiv \pi_0 + \alpha\pi_1 + \cdots$, $\bar{\sigma}_\alpha \equiv \bar{\sigma}_0 + \alpha\bar{\sigma}_1 + \cdots$. Inserting these expansions for the field variables and collecting all terms of order $\alpha^{-2}$ leads to the following Eikonal equation in matrix form.

$$\begin{bmatrix} -H_2 & 0^T \\ 0 & B \end{bmatrix}\begin{pmatrix} \pi_0 \\ \bar{\sigma}_0 \end{pmatrix} = 0 \tag{23}$$

where $H_2$ is defined in equations 8 and 9 and $B \equiv c^2\bar{\nabla}\varphi \otimes \bar{\nabla}\varphi - 1_{3\times 3}\left(D^*\varphi\right)^2$. Defining $Q \equiv -H_2 \oplus B$ for the block diagonal matrix in equation 23, the condition for the existence of a non trivial solution to equation 23, $\det Q = -H_2 \det B = 0$, leads to the condition

$$Q = \left(D^*\varphi\right)^2\left(\left(D^*\varphi\right)^2 - c^2\bar{\nabla}\varphi\cdot\bar{\nabla}\varphi\right) = 0. \tag{24}$$

Equations 21 and 22 represent a linear approximation which takes into account the interaction of acoustic waves with the background fluid motion and is quite a bit more involved than the results obtained by the relativity community, [5], [6]. Finally, by inspection of the terms $\widetilde{\Delta}_1$ and $\widetilde{\widetilde{\Delta}}_2$, one sees that they may be set to zero in the limit of a slowly varying environment since every term contains derivatives of the background fluid variables. This leads to a decoupling of the equations,

$$D^* \frac{1}{\rho_0 c^2} D^* p_1 - \vec{\nabla} \cdot \frac{1}{\rho_0} \vec{\nabla} p_1 \approx 0, \tag{25}$$

$$D^* \rho_0 D^* \vec{v} - \vec{\nabla} \rho_0 c^2 \vec{\nabla} \cdot \vec{v} \approx 0. \tag{26}$$

Most discussions on wave mechanics begin with a second order partial differential equation like equation 25. It is clear that the original equations must govern the behavior of the system in all situations.

In general geodesics are curves of "optimal" length, either minimizing or maximizing the length between two points of $M$. These curves may also be described as curves that parallel transport their velocity vector, referred to as auto-parallel. The condition for auto-parallelism on a differentiable manifold with a metric compatable torsion free connection is $T^\mu D_\mu T^\nu = 0$, where $T^\mu D_\mu A^\nu \equiv T^\mu \partial_\mu A^\nu + \Gamma^\nu{}_{\alpha\beta} T^\alpha A^\beta$ defines the covariant derivative of a contravariant vector $A^\nu$ along the direction $T^\mu$ and $2\Gamma^\alpha{}_{\mu\nu} \equiv g^{\alpha\beta}(\partial_\mu g_{\nu\beta} + \partial_\nu g_{\mu\beta} - \partial_\beta g_{\mu\nu})$ are the Christoffel symbols of the second kind. In order to ensure that the geodesic, or auto-parallel, condition hold the parameter, $\lambda$, defined along the curve must be chosen to be an affine parameter. If an other arbitrary parameterization is used the auto-parallel condition will generalize to $T^\mu D_\mu T^\nu = \sigma T^\nu$, where $\sigma$ is an arbitrary scalar function defined along the curve. This looser condition makes the transported tangent vector "parallel" to the tangent of the curve at the new point allowing a change in magnitude, whereas the auto-parallel condition requires the transported tangent vector to be identical to the tangent vector at the new point along the curve. The relation between affine parameterization and parallel transport are well discussed in the literature on differential geometry and general relativity. In particular for geodesics on Riemannian manifolds, manifolds that are locally Euclidian, arc length may always be used as an affine parameter. For Lorentzian manifolds, a generalization of Riemannian manifold where the space locally resembles Minkowski space introduced in special relativity, curves may have positive, negative or zero length. If the time direction is chosen to be negative (as in this paper) then curves with negative length are called time like, curves with positive length are called space like and curves with zero length are called null, or light like. Hence, space like curves may be arc length parameterized and each time like curve may be parameterized by a standard internal time, identified in general relativity as the proper time of an observer whose world line coincides with the particular time like curve. Clearly for null curves arc length cannot be used as a parameter. In such cases one imposes the strict form of the auto-parallel condition and the null constraint thus defining the parameter as affine with no physical significance attached.

The surface of constant time $\varphi(\bar{x},t)_{t=t_0} \equiv \varphi(\bar{x})$ generated by taking a constant time cross section of the four dimensional eikonal is referred to as a phase surface or wave front (from here on a wave front is a surface of equal time and "in phase" is taken to mean equal time of flight). The metric given in equation 11 leads to the null constraint $g_{00}dt^2 + g_{ij}dx^i dx^j + 2g_{0i}dtdx^i = 0$. From this null constraint one can derive the following constraint on $d\bar{x}/dt$

$$\frac{g_{ij}}{g_{00}}\frac{dx^i}{dt}\frac{dx^j}{dt} + 2\frac{g_{0i}}{g_{00}}\frac{dx^i}{dt} = -1. \tag{27}$$

For the metric given in equation 11, equation 27 states that the ray velocity relative to the moving fluid equals the local sound speed $(d\bar{x}/dt - \bar{v})\cdot(d\bar{x}/dt - \bar{v}) = c^2$ (see figure 1). An equation for $d\bar{x}/dt$ can be derived from equation 15 by writing the equations for time and space separately.

$$\frac{d^2 x^k}{d\lambda^2} + \Gamma^k{}_{ij}\frac{dx^i}{d\lambda}\frac{dx^j}{d\lambda} + \Gamma^k{}_{00}\frac{dt}{d\lambda}\frac{dt}{d\lambda} + 2\Gamma^k{}_{i0}\frac{dx^i}{d\lambda}\frac{dt}{d\lambda} = 0 \tag{28}$$

$$\frac{d^2 t}{d\lambda^2} + \Gamma^0{}_{ij}\frac{dx^i}{d\lambda}\frac{dx^j}{d\lambda} + \Gamma^0{}_{00}\frac{dt}{d\lambda}\frac{dt}{d\lambda} + 2\Gamma^0{}_{i0}\frac{dx^i}{d\lambda}\frac{dt}{d\lambda} = 0 \tag{29}$$

Changing the parameter along the ray from $\lambda$ to $t$ leads to,

$$\frac{d^2 x^k}{dt^2} + \frac{d^2 t/d\lambda^2}{(dt/d\lambda)^2}\frac{dx^k}{dt} + \Gamma^k{}_{ij}\frac{dx^i}{dt}\frac{dx^j}{dt} + \Gamma^k{}_{00} + 2\Gamma^k{}_{i0}\frac{dx^i}{dt} = 0 \tag{30}$$

$$\frac{d^2 t/d\lambda^2}{(dt/d\lambda)^2} + \Gamma^0{}_{ij}\frac{dx^i}{dt}\frac{dx^j}{dt} + \Gamma^0{}_{00} + 2\Gamma^0{}_{i0}\frac{dx^i}{dt} = 0 \tag{31}$$

Using equation 31 to replace the second term in equation 30 gives

$$\frac{d^2 x^k}{dt^2} - \left(\Gamma^0{}_{ij}\frac{dx^i}{dt}\frac{dx^j}{dt} + \Gamma^0{}_{00} + 2\Gamma^0{}_{i0}\frac{dx^i}{dt}\right)\frac{dx^k}{dt}$$
$$+ \Gamma^k{}_{ij}\frac{dx^i}{dt}\frac{dx^j}{dt} + \Gamma^k{}_{00} + 2\Gamma^k{}_{i0}\frac{dx^i}{dt} = 0 \tag{32}$$

Substitution of the metric in equation one leads to the following for the Christoffel symbols.

$$\Gamma^0{}_{00} = d_{-\bar{v}}\ln c + \frac{v^i v^j}{2c^2}S_{ij} \tag{33a}$$

$$\Gamma^0{}_{i0} = \partial_i \ln c - \frac{\upsilon^j}{2c^2} S_{ij} \tag{33b}$$

$$\Gamma^i{}_{00} = \upsilon^i d_{-\bar{\upsilon}} \ln c - \partial_t \upsilon^i + c\partial_i c - \frac{1}{2}\upsilon^j \omega_{ij} - \frac{\upsilon^j}{2c^2}\left(c^2 \delta_{ki} - \upsilon^k \upsilon^i\right) S_{jk} \tag{33c}$$

$$\Gamma^i{}_{j0} = \upsilon^i \partial_j \ln c + \frac{1}{2}\omega_{ij} - \frac{\upsilon^i \upsilon^k}{2c^2} S_{jk} \tag{33d}$$

$$\Gamma^k{}_{ij} = \upsilon^k \Gamma^0{}_{ij} = \frac{\upsilon^k}{2c^2} S_{ij} \tag{33e}$$

Expressions for the Christoffel symbols involving spatial derivatives of the fluid velocity have been reduced as far as possible to terms involving the quantities such as the fluid vorticity $\omega_{ij} \equiv \partial_i \upsilon_j - \partial_j \upsilon_i$ and fluid shear and compression tensor $S_{ij} \equiv \partial_i \upsilon_j + \partial_j \upsilon_i$. This reduction is natural from a physical point of view as effects due to fluid vorticity *etc* become obvious. Similar expression may be found for the Christoffel symbols of the first kind which are needed later in calculations of the covariant Riemann tensor.

Using the reduced Christoffel symbols and after some algebra the equation for the 3 dimensional ray paths as a function of time becomes

$$\ddot{x}^k = \frac{\left(\dot{x}^k - \upsilon^k\right)}{2c^2} S_{ij}\left(\dot{x}^i - \upsilon^i\right)\left(\dot{x}^j - \upsilon^j\right) + \left(\dot{x}^k - \upsilon^k\right)\left(d_{-\bar{\upsilon}} \ln c + 2\dot{\bar{r}} \cdot \bar{\nabla} \ln c\right)$$
$$+ \partial_t \upsilon^k - c\partial_k c - \omega_{kj}\left(\dot{x}^j - \upsilon^j\right) + \frac{1}{2}\left(S_{kj} - \omega_{kj}\right)\upsilon^j \tag{34}$$

where $\dot{\bar{r}} = \dot{x}^k \hat{e}_k$ and $d_{-\bar{\upsilon}} = \partial_t - \bar{\upsilon} \cdot \bar{\nabla}$, the directional derivative along $-\bar{\upsilon}$, are introduced simply as a convenient shorthand.

Equation 34 may be converted to an equation for the wave front gradient through the identification $c\hat{n} \equiv \dot{\bar{x}} - \bar{\upsilon}$ and can be used to determine either the evolution of $\bar{x}$ or $\hat{n}$. Various forms of equation 34 appear in the literature parameterized by the three dimensional arclength *s*. In arclength parameterization the ray equation takes on the standard form of a unit speed curve in $E^3$, allowing for the identification of ray curvature and torsion, concepts familiar to the study of regular curves embedded in $E^3$.

### II. 2  GEODESIC DEVIATION, AFFINE PARAMETERIZATION

Given a specific solution to equation 15, labeled $\gamma_F(\lambda)$ and called a fiducial ray, the behavior of neighboring geodesics is governed by the Jacobi equation,

$$\frac{D^2 Y^\mu}{d\lambda^2} + \left(R^\mu{}_{\alpha\nu\beta} T^\alpha T^\beta\right)\Big|_{\gamma(\lambda)} Y^\nu = 0 \tag{35}$$

where, $T^\alpha$ is the tangent to the affine parameterized geodesic $\gamma_F(\lambda)$. The specific solution $\gamma_F(\lambda)$ is used as a reference curve for tracing geodesics within a neighborhood of $\gamma_F(\lambda)$. The coordinates of the fiducial geodesic are labeled $x_F^\mu$ and the coordinates of the neighboring geodesic $x_N^\mu = x_F^\mu + Y^\mu$. The components of the Riemann curvature, $R^\mu{}_{\alpha\nu\beta}$, and the covariant Riemann tensor, $R_{\mu\alpha\nu\beta}$, are given in a coordinate basis by

$$R^\mu{}_{\alpha\nu\beta} = \partial_\nu \Gamma^\mu{}_{\alpha\beta} - \partial_\beta \Gamma^\mu{}_{\alpha\nu} + \Gamma^\mu{}_{\lambda\nu}\Gamma^\lambda{}_{\alpha\beta} - \Gamma^\mu{}_{\lambda\beta}\Gamma^\lambda{}_{\alpha\nu}, \tag{36}$$

$$R_{\mu\alpha\nu\beta} = \partial_\nu \{\mu;\alpha\beta\} - \partial_\beta \{\mu;\alpha\nu\} + \Gamma^\rho{}_{\alpha\nu}\{\rho;\mu\beta\} - \Gamma^\rho{}_{\alpha\beta}\{\rho;\mu\nu\}, \tag{37}$$

respectively. The Jacobi field $Y^\mu$, also referred to as the geodesic deviation vector, measures the local separation of neighboring geodesics relative to a chosen reference geodesic for equal values of $\lambda$. Equation 35 is a special case of a more general deviation equation for which $Y^\mu$ is constrained by $Y^\mu T^\nu g_{\mu\nu} = 0$, i.e. the deviation vector is held orthogonal to the geodesic [19], and is identical to the second variation of the action leading to equation 15. Consequently, this field determines both the stability of the solutions to the geodesic equation and provides estimates of solutions when the geodesic flow is determined to be stable. Equation 35 is converted into an ordinary differential equation in $\lambda$ by introducing a non rotating, pseudo-orthonormal, frame field along $\gamma_F(\lambda)$. Following the notation in Hawking *et al* [12], let $T^\mu$ serve as one of the coordinate directions then choose a second null vector, $L^\mu$, satisfying the condition $L^\mu T^\nu g_{\mu\nu} = -1$, as the second basis vector. To complete the local geodesic coordinate basis two space-like directions, $e_1^\alpha$ and $e_2^\beta$, satisfying the conditions $T^\alpha e_{I\alpha} = 0$, $L^\alpha e_{I\alpha} = 0$ and $e_I^\alpha e_{J\alpha} = \delta_{IJ}$ for $I, J = 1, 2$ are introduced. The two null vectors define a time-like direction $E_t = \frac{1}{c}(1\ \vec{v})$ that is orthogonal to the two-dimensional space-like hyper-surface $\langle \hat{e}_I \rangle$ at all points in space-time. After defining this pseudo-orthonormal basis at an initial point on $\gamma_F(\lambda)$ the basis is parallel transported along $\gamma_F(\lambda)$ to define a new basis at each point by solving the parallel transport equation.

$$T^\mu D_\mu (\hat{e}_I)^\alpha = \frac{d(\hat{e}_I)^\alpha}{d\lambda} + \Gamma^\alpha{}_{\mu\nu} T^\mu (\hat{e}_I)^\nu = 0 \tag{38}$$

An important consequence of this construction is that the deviation vector points in four-dimensional space-time from the fiducial geodesic, $\gamma_F(\lambda)$ to a point on a neighboring geodesic, $\gamma_N(\lambda)$, with the same value of affine parameter $\lambda$ and the deviation vector will not necessarily remain "in phase" with the fiducial ray path in the traditional sense of the term. This does not pose any problems in the geometric description of neighboring curves since the Jacobi field simply maps out the local space-time geometry in a tube surrounding $\gamma_F(\lambda)$ without prejudice to any coordinate, time being just another coordinate in the four dimensional geometric paradigm (see figure 2).

To ensure that neighboring rays are emitted from the source at the same time as the fiduciary ray the initial space-like basis is chosen to be purely spatial in the coordinate frame, $(\hat{e}_I)^\mu = (0 \quad \hat{e}_I)$ and the initial deviation $Y_0^\mu = (0 \quad \vec{Y}_0)$ is chosen to lie along one of the initial basis vectors. Even though the initial basis vectors are chosen such that $(\hat{e}_I)^\mu = (0 \quad \hat{e}_I)$ and $\hat{e}_I \cdot (\vec{t} - \vec{v}) = 0$, upon propagation along the geodesic $\gamma(\lambda)$ the vectors $\langle \hat{e}_I \rangle$ will pick up time components and the purely spatial portion will not in general remain parallel to the three dimensional surface of equal time (see figure 3).

Projected into the non rotating pseudo-orthonormal basis equation 35 is reduced to the linear second order equation,

$$\frac{d^2 Y_I}{d\lambda^2} + K_{IJ} Y_J = 0 \tag{39}$$

with, $Y_I = \hat{e}_I^\alpha g_{\alpha\beta} Y^\beta$, and the curvature matrix $K_{IJ} \equiv R_{\mu\alpha\nu\beta} T^\alpha T^\beta \hat{e}_I^\mu \hat{e}_J^\nu$ is introduced, sum over $J = 1, 2$ is implied in equation 39. The paraxial ray tracing procedure is now constructed by taking equations 15, 38 and 39 together as a single system.

The covariant Riemann tensor components in a Cartesian coordinate basis derived from the metric in equation 11 are listed below.

$$R_{nimj} = \frac{1}{4c^2}\left(S_{mn} S_{ij} - S_{mi} S_{nj}\right) \tag{40}$$

$$R_{0ikj} = \frac{1}{2}\left(\partial_j S_{ik} - \partial_k S_{ij}\right) + \frac{1}{2}\left(S_{ij}\partial_k \ln c - S_{ik}\partial_j \ln c\right)$$
$$+ \frac{v_n}{4c^2}\left(S_{ik} S_{jn} - S_{ij} S_{kn} + S_{ik}\omega_{jn} - S_{ij}\omega_{kn}\right) \tag{41}$$

$$R_{0i0j} = -\frac{1}{2}\partial_t S_{ij} - \frac{1}{2}v_n\left(\partial_j S_{in} + \partial_j \omega_{in}\right) + c\partial_i\partial_j c + \frac{1}{2}S_{ij}\partial_t \ln c$$
$$+ \frac{v_n v_m}{4c^2} S_{in} S_{jm} - \frac{1}{4}\left(S_{im} S_{jm} + \omega_{im} S_{jm} + \omega_{jm} S_{im}\right)$$
$$+ \frac{v_n}{2}\left(S_{jn}\partial_i \ln c + S_{in}\partial_j \ln c - S_{ij}\partial_n \ln c\right) \tag{42}$$

Other relevant terms may be calculated using the symmetries of the Riemann tensor, $R_{\mu\nu\alpha\beta} = R_{\alpha\beta\mu\nu} = -R_{\nu\mu\alpha\beta} = -R_{\mu\nu\beta\alpha}$.

### II. 3. GEODESIC DEVIATION, TIME PARAMETRIZATION

Using the metric given in equation 11 along with the null constraint it is a straight forward matter to show that the four dimensional orthogonality relation $T^\mu \hat{e}_I^\nu g_{\mu\nu} = 0$ implies the relation $(\vec{t} - \vec{v}) \cdot (\hat{e}_I - \hat{e}_I^0 \vec{t}) T^0 = 0$ and with a little algebra one can show that the normalization condition $\hat{e}_I^\mu \hat{e}_I^\nu g_{\mu\nu} = 1$ implies $\|\hat{e}_I - \vec{t}\hat{e}_I^0\| = 1$ with respect to the ordinary dot product in three dimensional space, where in both cases the null constraint is used. Hence the three dimensional vector $\tilde{e}^k \equiv \hat{e}_I^k - \hat{e}_I^0 \, dx^k/dt$ is tangent to the wave front everywhere along the ray, $\hat{n} \cdot \tilde{e} = 0$, and obeys the three dimensional orthonormality condition as it is propagated along the ray. (The set of vectors $\langle \tilde{e}_I \rangle$ is referred to here as an auxiliary basis.) Once the full deviation $Y^\mu = Y_1 \hat{e}_1^\mu + Y_2 \hat{e}_2^\mu$ is calculated the equal time deviation vector $\tilde{Y}^k = Y_1(\hat{e}_1^k - t^k \hat{e}_1^0) + Y_2(\hat{e}_2^k - t^k \hat{e}_2^0)$ may be constructed with the result that $\|\tilde{Y}\| = \sqrt{Y^\mu Y_\mu} = \sqrt{Y_1^2 + Y_2^2}$, (see figure 4).

Expanding equation 38 into a sum over individual components of the Riemann tensor only terms of the form $R^\mu{}_{\alpha\nu\beta} T^\alpha \Lambda_I{}^{\nu\beta}$ arise, where the second rank antisymmetric tensor $\Lambda_I{}^{\mu\nu} \equiv T^\mu \hat{e}_I^\nu - T^\nu \hat{e}_I^\mu$ is defined. The tensor, $\Lambda_I{}^{\mu\nu}$, is parallel transported along any geodesic, (this follows from the fact that $T^\mu$ and $\hat{e}_I^\nu$ are parallel transported that any higher order tensor constructed from them is parallel transported as well), and naturally describes the three dimensional wave front embedded in three dimensional Euclidian space. The mixed (time-space) components of the tensor, $\Lambda_I{}^{0k} = T^0 \hat{e}_I^k - T^k \hat{e}_I^0 = T^0(\hat{e}_I^k - (dx^k/dt)\hat{e}_I^0) = T^0 \tilde{e}_I^k$, are clearly proportional to the local tangent vectors on the wave front, *i.e.* the auxiliary basis defined earlier while the pure space components, $\Lambda_I{}^{ik} = T^i \hat{e}_I^k - T^k \hat{e}_I^i = T^i \tilde{e}_I^k - T^k \tilde{e}_I^i$, are equal to the components of the cross product $\vec{T} \times \tilde{e}_I$ in three dimensions. The transport equation for $\Lambda_I{}^{\mu\nu}$ is,

$$\frac{d}{d\lambda} \Lambda_I{}^{\mu\nu} + \Gamma^\mu{}_{\alpha\beta} T^\alpha \Lambda_I{}^{\beta\nu} + \Gamma^\nu{}_{\alpha\beta} T^\alpha \Lambda_I{}^{\mu\beta} = 0. \tag{43}$$

Either of equations 38 or 43 may be used to determine the basis along the ray. When the free index appearing in the deviation equation, equation 35, is summed over the internal basis the second terms reduces to $4K_{IJ} = R_{\mu\nu\alpha\beta} \Lambda_I{}^{\mu\nu} \Lambda_J{}^{\alpha\beta}$.

The auxiliary basis naturally frames the wave front and is a generalization of the wave front basis already employed in paraxial techniques used in Seismology. Equation 43 tracks six fields, counting indices in four dimensions. It is clear from the definition of $\Lambda_I{}^{\mu\nu}$ that there are only three degrees of freedom in equation 43. By separating equation 43 into individual components the following transport equation for the wave front basis $\tilde{e}_I^k$ in $E^3$ may be derived

$$\frac{d}{dt} \tilde{e}_I^k + \frac{1}{2} \{\omega_{ki} - n_k n_j S_{ji} - 2n_k \partial_i c\} \tilde{e}_I^i = 0 \tag{46}$$

for the wave front basis as a function of time evaluated along the ray in $E^3$. By symmetry the second term in 46 may be written

$$\Omega_{ki} \equiv \frac{1}{2}\varepsilon_{jki}\left(\vec{\nabla}\times\vec{\upsilon}+\vec{s}\times\hat{n}+2\vec{\nabla}c\times\hat{n}\right)_j, \tag{47}$$

with $s_k \equiv n_j S_{jk}$, which makes the identification of equation 46 with an infinitesimal three dimensional rotation in $E^3$ apparent. A similar first order equation for the evolution of the wave front normal may be derived from the ray equation using the definition $\vec{t} = c\hat{n} + \vec{\upsilon}$,

$$\frac{d}{dt}\hat{n}_k + \Omega_{ki}\hat{n}_i = 0. \tag{48}$$

In principle all transport equations have solutions in terms of path ordered integrals involving the SO(3) generator evaluated along the ray. In practice these are not known in closed form. However assuming that the ray equation has been solved either analytically or numerically the normal $\hat{n}$ is known allowing one to reduce the problem to a two dimensional rotation about the wave front normal followed by an application two Euler matrices to orient the basis along the ray path.

Choosing the initial basis to match a global Cartesian coordinate basis $\langle \hat{n}(0) \; \tilde{e}_1(0) \; \tilde{e}_2(0)\rangle = \langle i \; j \; k\rangle$ the rotation matrix taking $\hat{n}(0) \to \hat{n}$ may be found explicitly.

The action of this sequence on the initial basis vectors, interpreted as an active transformation on the initial values $\langle \hat{n}(0) \; \tilde{e}_I(0)\rangle$, is denoted $U = R(\varphi)R(\theta)R(\alpha)$. The rotation $R(\alpha)$ has a block diagonal form $R(\alpha) = 1 \oplus r$, where $r$ is a two dimensional rotation about the $x$–axis of the global coordinate system. Equation 49 leads to the following differential equation for $r$.

$$\frac{dr}{dt} + \tilde{\omega}\sigma_x r = 0 \tag{50}$$

where $\sigma_x = \begin{pmatrix} 0 & -1 \\ 1 & 0 \end{pmatrix}$ and $\tilde{\omega} \equiv \dfrac{n_y\Omega_{xz} - n_x\Omega_{yz}}{n_x^2 + n_y^2}$. A closed form solution to equation 50 may be found,

$$r = \begin{pmatrix} \cos\alpha & -\sin\alpha \\ \sin\alpha & \cos\alpha \end{pmatrix} \tag{51}$$

where $\alpha = \int \tilde{\omega}\big|_\gamma dt$ with the integrand evaluated along the specific ray path in space, evaluation of the solution requires a full solution to the ray equation.

Changing variables from $\lambda$ to $t$ in the first term of equation 39, recognizing that the second term may be written $K_{IJ} = (T^0)^2 R_{\mu\alpha\nu\beta}t^\alpha t^\beta \hat{e}_I^\mu \hat{e}_J^\nu \equiv (T^0)^2 \tilde{K}_{IJ}$ and using equation 31 leads to the following equation for the deviation vector along $\gamma_F(t)$ as a function of coordinate time

$$\frac{d^2 Y_J}{dt^2} - \kappa \frac{dY_J}{dt} + \widetilde{K}_{IJ} Y_I = 0, \qquad (52)$$

with $\kappa \equiv \Gamma^0{}_{mn} t^m t^n + \Gamma^0{}_{00} + 2\Gamma^0{}_{m0} t^m$ and $\widetilde{K}_{IJ} = \left(R_{\mu 0 \nu 0} + R_{\mu m \nu n} t^m t^n + 2 R_{\mu 0 \nu n} t^n \right) \hat{e}_I^\mu \hat{e}_J^\nu$. It is understood that the expressions in equation 52 that involve the Christoffel symbols and the Riemann tensor are evaluated along the ray path $\gamma_F(t)$, being expressed as functions of time. The paraxial ray tracing system introduced in the last section may now be cast in a more tractable form by taking equations 34, 46 and 52 to form the system of paraxial ray equations. The deviation vector produced by this procedure tracks rays that are in phase with respect to the central ray in the system and tracks them all in true coordinate time.

## II.4 INITIAL CONDITIONS, MAPPING OF NIEGHBORING RAYS AND BEAM DEFORMATION

The components of the equal time deviation vector $\widetilde{Y}_I$ approximate a small element of arc length within the Eikonal surface in the $\widetilde{e}_I$ direction. The initial value of the deviation may be chosen arbitrarily. The initial value of $\dot{Y}_I$ may also be chosen arbitrarily and is related to the values of the physical parameters that describe the environment. In modeling the evolution of a wave front the initial geometry is assumed known, hence it is desirable to express $\dot{Y}_{I0}$ in terms of known quantities. By definition $x_N^\mu = x_F^\mu + Y^\mu$, $\dot{Y}^\mu = \dot{x}_N^\mu - \dot{x}_F^\mu$ and $d\bar{x}_F / d\lambda = dt_F / d\lambda \left( c_F \hat{n}_F + \vec{v}_F \right)$ with a similar expression for the neighboring ray, $c_F \equiv c(x_F^\mu)$ being a short hand notation for the local sound speed evaluated at a point on the fiduciary ray (with similar expressions for all other quantities evaluated along the ray). At the initial wave front $t_F|_0 = t_N|_0$ by choice and the initial rates may chosen such that $dt_F / d\lambda|_0 = dt_N / d\lambda|_0$. Denoting this common initial rate of the time coordinate by the constant $\beta$ the initial velocity of the deviation vector becomes $\dot{\vec{Y}}_0 = \beta \left( c_N \hat{n}_N - c_F \hat{n}_F + \vec{v}_N - \vec{v}_F \right)$, all quantities being evaluated at the same $\lambda$ but along different rays. In terms of time parameterization $d\vec{Y}/dt|_0 = c_N \hat{n}_N - c_F \hat{n}_F + \vec{v}_N - \vec{v}_F$, and the initial deviation velocity for two rays with a common initial position $d\vec{Y}/dt|_0 = c_0 \Delta \hat{n}$, where $\Delta \hat{n} \equiv \hat{n}_N - \hat{n}_F$ with both terms being evaluated at the same point in space. The quantity $\Delta \hat{n}$ determines the initial shape of the wave front whereas all other quantities are assumed to be given (see figure 5).

The initial values of $\dot{Y}_I$ may be expressed in terms of $\dot{Y}_0^\mu$. By explicitly differentiating the expression $Y_I = Y^\mu \hat{e}_{I\mu}$ making use of equation 38, the explicit for of the Christoffel symbols, and the results of the last paragraph the following expressions for the initial values of $\dot{Y}_I$ are derived,

$$\left.\frac{dY_I}{d\lambda}\right|_0 = \beta \hat{e}_I^k\Big|_0 \left\{\frac{dY^k}{d\lambda} + \frac{1}{2}Y^i\omega_{ki}\right\}\Big|_0 \qquad (53)$$

where all values are taken at the initial point along $\gamma_F(\lambda)$ and $dY^k/d\lambda\big|_0$ is given in the previous paragraph in terms of data at the initial point along both $\gamma_F(\lambda)$ and $\gamma_N(\lambda)$. Dividing by $\beta$ in equation 46 gives the appropriate initial values for $dY_I/dt$. Given a specific choice of initial wave front geometry at a specific position in space time equation 41 gives the appropriate initial conditions for $dY_I/d\lambda$ or $dY_I/dt$.

Two points, P and Q, along a geodesic $\gamma(\lambda)$ are said to be conjugate if there exists a nontrivial Jacobi vector along $\gamma(\lambda)$ that vanishes at both P and Q. Intuitively caustics, or focal points, along a ray may be identified with points where $Y^\mu = 0$. To be more explicit the deviation vector is expressed as a function of the initial conditions $Y(\lambda;(Y_0,\dot{Y}_0)) \equiv Y_\lambda(a_i)$, where $(Y_0,\dot{Y}_0) \equiv a_i$ is a short hand notation referring to the set of initial conditions along the $i$-th neighboring ray to $\gamma_F(\lambda)$ and coordinate indices are suppressed. If the initial conditions of the family of neighboring rays are close to those of the fiduciary ray (which is necessarily true) then a caustic point along $\gamma_F(\lambda)$ is determined by $Y_\lambda(a_i) = 0$. It may happen that two different neighboring rays, say $\gamma_{N_1}(\lambda)$ and $\gamma_{N_2}(\lambda)$, with similar initial conditions meet without ever encountering $\gamma_F(\lambda)$. If the two sets of initial conditions are close to each other a caustic is determined by the condition $Y_\lambda(a_1) = Y_\lambda(a_2)$.

In general conjugate points will exist when the curvature in the $\langle T^\mu, Y^\mu \rangle$ plane is positive definite. This can be seen by projecting the components of Riemann into the $\langle T^\mu, Y^\mu \rangle$ plane which leads to the standard quadratic form for the sectional curvature $K_{IJ}Y_IY_J$. If the curvature is positive definite in every plane containing the ray tangent then neighboring rays from a common source point in all directions tend to converge and a three dimensional ray bundle will focus, forming caustics periodically. If the curvature is zero or negative everywhere along the ray then neighboring rays from a common source point will tend to eventually diverge. (If $Y_0 \neq 0$ then the initial shape of the wave front may cause the formation of a caustic independently of the focusing properties of the medium, as happens with light reflected from a concave surface in Euclidian space). The conditions $K_{11} > 0$, $K_{22} > 0$ and $\det K_{IJ} > 0$ along $\gamma(\lambda)$ are necessary and sufficient for the quadratic form, $K_{IJ}Y_IY_J$, to be positive definite.

Applying the conjugate point theorems of differential geometry [13] if upper and lower bounds exist such that $k_1 \leq R(\hat{Y},T,\hat{Y},T) \leq k_2$ along some segment of the ray then the period of affine parameter between consecutive conjugate points satisfies $\pi/\sqrt{k_2} \leq T \leq \pi/\sqrt{k_1}$. If the sectional curvature is zero or negative everywhere along $\gamma_F(\lambda)$ then neighboring rays will eventually diverge (a single caustic may form if the initial region is concave but otherwise one does not expect the periodic development of focal points further down stream). In situations where a ray may enter convergence and divergence zones occasionally one can not conclude from checking the Riemann tensor whether conjugate points will occur and how frequently.

By tracing $\widetilde{Y}(t, a_i)$ for fixed $t$ and all values of initial conditions, a picture of the beam cross section at any later time may be constructed. The area of the planar cross section is calculated using $A(t) = \frac{1}{2}\left|\oint \widetilde{Y}(t) \times d\widetilde{Y}(t)\right|$, where $\widetilde{Y}$ and $d\widetilde{Y}$ may be expressed in terms of $a_i$ and $da_i$ (see figure 6). When the acoustic energy contained within a ray bundle remains constant the acoustic intensity at any point along the ray in inversely proportional to the cross sectional area of the ray bundle. Defining the differential area element $\widetilde{Y} \times d\widetilde{Y} = dA\hat{n}$ leads to $I_0/I_t = (dA\hat{n}\cdot\hat{t})_t/(dA\hat{n}\cdot\hat{t})_0$, where $\hat{t}$ is a unit vector tangent to the ray path.

## III  APPLICATION TO LAYERED MEDIA, GENERAL SOLUTION

To find an explicit expression for the deviation vector the tangent of the fiduciary ray and the basis vectors along the ray must be known at every point. For two dimensional problems involving layered media it is always possible to find closed form expressions for these quantities thus making the reduction of equation 39 to an ordinary linear differential equation in one variable possible. To achieve this, the explicit form of the geodesic equation, equation 15 is abandoned in favor of a set of first order equations derived from identifying isometries of the metric tensor.

The presence of a cyclic coordinate in the metric, labeled $x^C$, leads to a conservation law for the corresponding momentum or conjugate variable,

$$\frac{dp_C}{d\lambda} = \frac{dg_{C\beta}T^\beta}{d\lambda} = \frac{d\xi_C^\alpha g_{\alpha\beta}T^\beta}{d\lambda} = 0 \tag{54}$$

where $\xi_C^\alpha$ is the Killing vector in the direction of the $x^C$ coordinate curves [15]. (The statement "…presence of a cyclic coordinate, $x^C$" is equivalent to the absence of that coordinate from the metric altogether).

The ray equations are derived in three dimensions then reduced to the two dimensional case later. Specializing to a time independent environment with Killing vector $\xi_t^\alpha = \begin{pmatrix} 1 & \vec{0} \end{pmatrix}$ leads to the conservation law

$$-(c^2 - \upsilon^2)\frac{dt}{d\lambda} - \vec{\upsilon}\cdot\frac{d\vec{r}}{d\lambda} = \kappa_0. \tag{55}$$

Traditionally this is associated with the energy of the particle and for the metric signature in equation 11 $\kappa_0$ is negative (from here on $\kappa_0 > 0$ and an overall minus sign is dropped from equation 55). Next, consider the problem of a layered medium where $c$ and $\vec{\upsilon}$ depend on only one coordinate, say $z$. There are two more conserved currents, which may be thought of as components of the ray's translational momentum. With $x$ the second cyclic coordinate the Killing vector $\xi_x^\alpha = (0 \ 1 \ 0 \ 0)$ leads to the corresponding conservation law,

$$\frac{dx}{d\lambda} - \frac{dt}{d\lambda}v^x = \kappa_1, \tag{56}$$

and an identical expression existing for the $y$ component of momentum with all occurrences of $x$ changed to $y$ in equation 56 along with the definition of the conserved current $\kappa_2$. Imposing the null constraint on the components of the tangent vector leads to,

$$-c^2\dot{t}^2 + \vec{\kappa}\cdot\vec{\kappa} + (\dot{z} - v_z\dot{t})^2 = 0, \tag{57}$$

where the following notation has been introduced, $\vec{\kappa} \equiv \kappa_1\hat{i} + \kappa_2\hat{j}$ and $\vec{v}_T \equiv v_x\hat{i} + v_y\hat{j}$ are both $E^3$ vectors. Equation 57 may be further reduced by applying equation 55 to give

$$c^2\dot{t} = \kappa_0 - \vec{v}_T\cdot\vec{\kappa} - v_z(\dot{z} - v_z\dot{t}) \tag{58}$$

Equations 57 and 58 yield the following differential equations for the depth and travel time

$$\dot{z} = \pm\frac{1}{c}\sqrt{(\kappa_0 - \vec{v}_T\cdot\vec{\kappa})^2 - \kappa^2 c^2(1-\beta_z^2)} \tag{59}$$

$$\dot{t} = \frac{\kappa_0 - \vec{v}_T\cdot\vec{\kappa} \mp \beta_z\sqrt{(\kappa_0 - \vec{v}_T\cdot\vec{\kappa})^2 - \kappa^2 c^2(1-\beta_z^2)}}{c^2(1-\beta_z^2)} \tag{60}$$

where $\beta_z \equiv v_z/c$, the vertical rapidity, has been introduced. Equation 60 may then be inserted into equation 57 to yield a set of first order differential relations for $\dot{x}^\mu(z)$.

Consider a medium with one dimensional fluid velocity of the form $v_x(z) \equiv v(z)$. Rays, considered as curves in $E^3$, that are initially fired in the $x - z$ plane do not turn out of their initial osculating plane, *i.e.* are torsion free, and constitute an effective two dimensional system (this follows from the fact that $dy/d\lambda = \kappa_2 = 0 \Rightarrow d^2y/d\lambda^2 = 0$).

Equations 48 through 50 lead to the following first order ordinary differential equations for the ray trajectory.

$$\frac{dt}{d\lambda} = \frac{\kappa_0}{c^2}(1 - v\alpha) \tag{63}$$

$$\frac{dx}{d\lambda} = \kappa_0 \alpha + \kappa_0 \frac{\upsilon}{c^2}(1 - \upsilon\alpha) \tag{64}$$

$$\frac{dz}{d\lambda} = \pm \frac{\kappa_0}{c}\sqrt{(1-\upsilon\alpha)^2 - \alpha^2 c^2} \tag{65}$$

Differentiating equation 63 implies $c^2 \ddot{t} = -(\kappa_1 \upsilon' + 2cc'\dot{t})\dot{z}$ while differentiation of equation 65 and use of the last result implies $\ddot{z} = -(\kappa_1 \upsilon' + cc'\dot{t})\dot{t}$, where dot denotes differentiation with respect to $\lambda$. It may be demonstrated that turning points in time are not allowed, as the relation $dt/d\lambda = 0$ leads to the unphysical result $(dz/d\lambda)^2 < 0$, hence without loss of generality the condition $dt/d\lambda > 0$ may be placed on any ray in the system.

Equation 65 may be rearranged, $c^2 (\dot{z})^2 / \kappa_1^2 = (\alpha - (\upsilon + c))(\alpha - (\upsilon - c))$, and the ray parameter determined from the equations of motion $\alpha \equiv \cos\theta_0 / (c_0 + \upsilon_0 \cos\theta_0)$, where $c_0$ and $\upsilon_0$ are the sound speed and fluid velocity at the initial position and $\theta_0$ is the initial angle between the wave front normal and the $x$-axis.

The single basis vector for the eikonal, $\tilde{e}$, may be constructed by simply rotating the normal $\hat{n} = d\bar{r}/dt - \bar{\upsilon}$ by $\pm 90°$ in the $x - z$ giving the following expressions for the components of the second rank antisymmetric tensor $\Lambda^{\mu\nu}$, $\hat{e}^x \dot{t} - \hat{e}^0 \dot{x} = \pm \dot{z}/c$ and $\hat{e}^z \dot{t} - \hat{e}^0 \dot{z} = \mp \kappa_1 / c$ and the useful relation $\hat{e}^x \dot{z} - \hat{e}^z \dot{x} = \pm \kappa_0 / c$, which is derived from the identity $T^i \hat{e}_I^k - T^k \hat{e}_I^i = T^i \tilde{e}_I^k - T^k \tilde{e}_I^i$ the expressions for the auxiliary basis and the ray equations 63 through 65. Thus the explicit form of the Christoffel symbols is unnecessary for determining the ray and its basis.

The Cartesian components of the covariant Riemann tensor derived from the acoustic metric and equation 37 are listed below.

$$R_{0x0x} = -c^2 R_{xzxz} = \frac{(\upsilon')^2}{4} \tag{66a}$$

$$R_{0zxz} = \frac{1}{4c^2}\left(2c^2 \upsilon'' + \upsilon(\upsilon')^2 - 2cc'\upsilon'\right) \tag{66b}$$

$$R_{0z0z} = -\frac{1}{4c^2}\left(-4c^3 c'' + 3c^2 (\upsilon')^2 + 4c^2 \upsilon\upsilon'' - 4cc'\upsilon\upsilon' + \upsilon^2 (\upsilon')^2\right) \tag{66c}$$

Using the basis vector $\hat{e}^\mu$ and the tangent $T^\mu = \dot{x}^\mu$ along $\gamma(\lambda)$ the sectional curvature may be calculated by expanding $K = R_{\mu\nu\alpha\beta}\Lambda^{\mu\nu}\Lambda^{\alpha\beta}/4$ explicitly to give

$$K = \frac{\kappa_0^2}{2}\frac{1}{c}\frac{d}{dz}\left(\frac{1}{c}\frac{d}{dz}\left(\alpha^2 c^2 - (1-\alpha\upsilon)^2\right)\right) \tag{68}$$

for the sectional curvature. The equation for the deviation vector reduces to a second order linear equation for a single scalar field, equation 69

$$\frac{d^2Y}{d\lambda^2} + \kappa_0^2 \left\{ \frac{\alpha}{c^2} \upsilon''(1-\alpha\upsilon) - \frac{\alpha^2}{c^2}(\upsilon')^2 - \frac{\alpha}{c^3} c'\upsilon'(1-\alpha\upsilon) + \frac{\alpha^2}{c} c'' \right\} Y = 0. \qquad (69)$$

The last term simply states that $c''$ governs the focusing properties of a stationary medium with an inhomogeneous sound speed. A ray propagating in a region with $c'' > 0$ will eventually encounter conjugate points while rays propagating in regions where $c'' < 0$ will diverge from one another. When the sound speed is constant the effects of fluid motion can be broken up into two terms, $\kappa_1 \dot{t} \upsilon'' - (\upsilon')^2 (\kappa_1/c)^2$. The second term is always negative, causing ray divergence. In the first term $\dot{t} > 0$ and the focusing effects are determined by the relative sign of $\kappa_1$ and $\upsilon''$. This term will cause focusing of acoustic rays when $\kappa_1$ and $\upsilon''$ are the same sign. As a consequence a constant fluid velocity has no effect on the stability of neighboring rays while a fluid velocity with a constant gradient will interfere with the focusing caused by an inhomogeneous sound speed with $c'' > 0$.

A more interesting situation occurs when both $c$ and $\upsilon$ depend on depth. The remarks of the last paragraph still hold true with the addition of an extra term coupling the sound speed gradient to the fluid velocity gradient, $-\upsilon' c' \kappa_1 \dot{t}/c$, appearing in equation 69. This term is more difficult to interpret than the others contributing the curvature. Consider a simple situation in which a wave guide is created by a sound speed profile with $c'' > 0$ everywhere. Above the waveguide axis $c' > 0$ while below the waveguide axis $c' < 0$. If the fluid motion is to the right and characterized by a uniform gradient then $\upsilon' > 0$. For acoustic rays fired to the right $\kappa_1 > 0$ resulting in a separation of neighboring rays above the wave guide axis and an enhanced convergence of rays below the waveguide axis. When the background fluid motion is weak and slowly varying the leading order terms in equation 68 are $\kappa_0 \kappa_1 \upsilon''/c^2 + \kappa_1^2 c''/c$ indicating that the dominant effects are due to the concavity of the environmental parameters.

Between turning points, $\dot{z} \neq 0$, equation 69 may be converted into a differential equation in the variable $z$ which in general reduces equation 69 to a single integral. Defining the variable $d\sigma = cdz$ and making use of equation 65 the sectional curvature may be written in a simple form $2K = -d^2(c\dot{z})^2/d\sigma^2$. Changing variables in the first term of equation 69 leads to the following general solution for the Jacobi field along any segment of the ray between consecutive turning points.

$$Y = \dot{z}c \left\{ c_1 \int \frac{cdz}{\dot{z}^3 c^3} + c_2 \right\}. \qquad (70)$$

With constants $c_1$ and $c_2$ set to match the initial conditions of the deviation vector equation 70 may be integrated to yield a solution for $Y$ along any segment of $\gamma_F(z)$ between consecutive turning points. For cases when $\dot{z} = 0$ identically along the ray one cannot perform the change of variables from $\lambda$ to $z$ and such cases must be treated separately. Equation 70 can be integrated in some cases to yield a closed form solution and may always be estimated by numerical integration. For the case of layered stationary media equation 70 becomes

$$Y = \sqrt{1-\alpha^2 c^2}\left\{c_1 \int \frac{c\,dz}{(1-\alpha^2 c^2)^{3/2}} + c_2\right\}.$$

Rays with $\alpha = 0$ have special significance in the linear case, acting as asymptotes for the other rays in the system [Kornhauser]. When $\alpha = 0$ equations 63 and 65 may be solved by inspection to yield $t = \kappa_0 \lambda/c$ and $z = \pm ct + z_0$. Equation 64 reduces to $\dot{x} = \dot{t}\upsilon(z)$, which states $dx/dt = \upsilon(z)$ along the ray. By a change of variables $x(z)$ $dx = \pm \mu(z) dz$, which may be integrated to yield $x = x_0 \pm \int \mu(z) dz$. These results generalize the special case considered by Kornhauser, which predicts for the linear case a special ray $x \sim z^2$ that acts as an asymptote for rays that do not have turning points in the $z$ direction. Finally, equation 68 reduces to $K = 0$ along the ray $\alpha = 0$, hence the Jacobi field for this case reduces to $Y = \dot{Y}_0 \lambda + Y_0$.

### III.2 LINEAR VELOCITY PROFILE

For a linear fluid velocity profile, $\upsilon = cz/L$ and $c =$ constant, the second term of equation 68 vanishes and $K = -\kappa_1^2/L^2$. The deviation equation may be integrated to yield $Y = (\dot{Y}_0/\sqrt{K})\sinh(\sqrt{K}\lambda) + Y_0 \cosh(\sqrt{K}\lambda)$. Hence neighboring rays with the same initial position will diverge at a rate of $Y = (\dot{Y}_0/\sqrt{K})\sinh(a\kappa_1\lambda/c)$ with respect to the affine parameter. The ray coordinates may also be solved in this case. Defining a new dimensionless variables along the ray $\zeta \equiv \alpha(cz/L)-1$, $\tau \equiv c^2\alpha t/L$, $\xi \equiv x/L$, $\sigma \equiv \kappa_1\lambda/L$ and the new ray parameter $\beta \equiv c\alpha$ equations 63 through 65 may be integrated to yield

$$\zeta = \beta \cosh(\sigma - \varphi) \tag{71}$$

$$\tau = \beta \sinh(\sigma - \varphi) - A_0 \tag{72}$$

$$\xi = \xi_0 + \frac{1}{2}\ln A_0 + \frac{1}{2}\sigma - \frac{1}{\beta}(2 - \beta\cosh(\sigma - \varphi))\sinh(\sigma - \varphi) \tag{73}$$

where the parameters $A_0 \equiv \sqrt{\zeta_0^2 - \beta^2} + \zeta_0$ and $\varphi \equiv \ln(\beta/A_0)$ have been defined. Equations 71 through 73 represent the affine parameterized coordinates for a segment of the ray with $\dot{z} > 0$ and the initial conditions are set so that $\tau(0) = 0$.

### III.3 WIND INDUCED SOUND DUCT

The effects of $\upsilon''$ on a ray system are illustrated for a class of velocity functions of the form $\upsilon(z) = c_0(z/L)^{2n}$, where $c_0$ is the local sound speed, assumed to be constant, $L$ a length scale chosen such that the flow becomes supersonic for $|z| > L$ and $n > 0$ is a positive integer. This class of functions have the following common properties: $\upsilon(z)$ is a monotonic increasing function for $z > 0$, $\upsilon(z) \geq 0$ for all $z$, $\upsilon'(z) > 0$ for $z > 0$ and $\upsilon''(z) > 0$ for all $z$ and that $\upsilon(-z) = \upsilon(z)$. For the time being attention will be paid to rays that remain in regions of subsonic flow.

From Snell's Law $\varepsilon(z) + \sec\theta = \varepsilon_0 + \sec\theta_0$, where $\varepsilon_0$ and $\theta_0$ are the Mach number and initial angle of inclination at the source point. For purposes of illustration it is assumed that the source is place at $z = 0$ and attention focused on rays fired in the upper half plane. For rays fired against the wind $\theta_0 > \pi/2$ and $\sec\theta_0 < -1$. Combined with Snell's law this information implies $\sec\theta < 0$ at all points along the ray indicating that rays with negative ray parameter will never turn back down toward the source. On the other hand rays with positive ray parameter will always encounter vertical turning points and because of the symmetry of $\upsilon(z)$ this will happen periodically.

The vertical turning points may be found from Snell's law by setting $\theta = 0$ and solving for $z$. Although the form of Snell's law used here is expressed in terms of the direction of the wave front normal the geometry of the rays indicates that $\theta = 0$ when the ray encounters a vertical turning point. The equation for the turning points becomes $\varepsilon(z) = -1 + \sec\theta_0$. The condition $2 > \sec\theta_0 \Rightarrow \theta_0 < \pi/3$ is imposed to single out rays that turn before entering supersonic regions, a conditions that holds for any choice of $\upsilon$. From equation 68 the sectional curvature is calculated

$$K = \frac{\kappa_0^2}{L^2 c_0^2} 2n(4n-1)\zeta^{2n-2} \cos\theta_0 \left(\frac{\cos\theta_0}{\cos\theta} - \frac{2n}{4n-1}\right) \tag{74}$$

for all rays in the system, where the dimensionless variable $\zeta \equiv z/L$ is used. Clearly the sign of $K$ depends on two factors, $\cos\theta_0$ and the term in parenthesis. Equation 74 will change sign when $\cos\theta = \cos\theta_0(2 - 1/2n)$, the same result holding for rays fired into and against the wind. Since $\cos\theta$ is an increasing function along rays with $\cos\theta_0 < 0$ the condition is never satisfied and these rays will remain in a divergence zone forever. Rays with $\cos\theta_0 > 0$ are a little more interesting. The curvature will remain positive along these rays only if $\cos\theta < \cos\theta_0(2 - 1/2n)$. Since $\upsilon$ is strictly increasing $\cos\theta$ is strictly increasing as well and it suffices to check the curvature at the turning point where $\cos\theta = 1$. The curvature will be zero at the turning point if $\cos\theta_0 = 2n/(4n-1)$. For $n = 1$ this leads to the constraint $\cos\theta_0 = 2/3$, or $\theta_0 \approx 48°$. Rays fired into the wind with $\cos\theta_0 > 2/3$ will remain in a region of positive curvature as they propagate down stream. Rays with $\cos\theta_0 < 2/3$ will begin with $K > 0$ and eventually along the ray $K \leq 0$ causing neighboring rays to begin to diverge with respect to $\lambda$, in general when $\cos\theta = 2n\cos\theta_0 /(4n-1)$ (figure 7).

In the case $n = 1$ the ray equations and the deviation equation may all be integrated between turning points to yield solutions for the time of flight, range and deviation vector as functions of depth in terms of elliptic functions.

## IV. STATIONARY MEDIA

The geometry of the null hyper - surface is immune to conformal changes in the metric tensor [15] as long as the affine parameter is appropriately changed, ($d\tilde{\lambda} = af^2 d\lambda$ and $\tilde{g}_{\alpha\beta} = f^2 g_{\alpha\beta}$, with $a$ = constant and $f = f(x^\mu)$ a scalar function). When $\bar{\upsilon}(\bar{x},t) = 0$ the metric appearing in equation 11 may be replaced by $ds^2 = -dt^2 + (d\bar{x}\cdot d\bar{x})/c^2(\bar{x},t)$ after performing a conformal transformation. Additionally if $c(\bar{x},t) = c(\bar{x})$ coordinate time becomes a proper affine parameter for the geodesics of the conformal metric and the null geodesics in space-time correspond to the space-like geodesics of a three dimensional space with conformal inner product [14] ($ds^2 = 0 \Rightarrow dt^2 = d\bar{x}\cdot d\bar{x}/c^2(\bar{x})$). In this new metric space "distance" is equivalent to the travel time between two points on the ray. The corresponding Chistoffel symbols and Riemann tensor are

$$\Gamma^k{}_{ij} = -\left(\delta_{kj}\partial_i \ln c + \delta_{ki}\partial_j \ln c - \delta_{ij}\partial_k \ln c\right), \tag{80}$$

$$R^k{}_{mjn} = \frac{1}{c^2}\left\{\Xi_{kjmnpq}c\partial_p\partial_q c + \left(\delta_{kn}\delta_{mj} - \delta_{kj}\delta_{mn}\right)\bar{\nabla}c\cdot\bar{\nabla}c\right\} \tag{81}$$

respectively, with $\Xi_{kjmnpq} \equiv \left\{\left(\delta_{kj}\delta_{pm} - \delta_{mj}\delta_{pk}\right)\delta_{qn} + \left(\delta_{mn}\delta_{qk} - \delta_{kn}\delta_{qm}\right)\delta_{pj}\right\}$. Letting $\dot{\bar{r}} = c\hat{n}$, $\hat{n}\cdot\hat{n} = 1$, dot referring here to time derivative, equation 34 reduces to the much simpler form $\dot{\hat{n}} = \hat{n}\times\left(\hat{n}\times\bar{\nabla}c\right)$ for the time evolution of the direction of the tangent vector and the resulting ray equation reduces to the standard form found in Landau and Lifshitz [24] (this was also derived here in section II.3.a).

A straight forward application of equation 38 to this case demonstrates that the equation for the parallel propagated frame field appearing in Cerveny [1] is identical to that presented here. The identification of equation 81 with the results found in Cerveny for the dynamical equations is less apparent. A significant difference in the approach taken here is the application of the conformal transformation to the null hypersurface. In moving to the time parameterization or arclength parameterization, Cerveny and others treat this as an ordinary parameter change void of any geometric meaning thus altering the paraxial equation by changing its form, introducing terms involving the first derivative of the deviation vector. By treating this parameter change as part of a conformal transformation the form of each equation is identical in passing from one parameterization to another. All things considered the identification of the deviation equation presented here with the dynamical equations appearing in Cerveny requires a more subtle approach than that taken for demonstrating the equivalence of the ray centered basis. To demonstrate the equivalence of the two approaches one must be sure that the results from this paper are compared to those of Cerveny in the appropriate parameterization, $d\lambda$. Projecting equation 68 into local ray centered coordinates leads to equation 4.1.70 of Cerveny, after a bit of algebra.

For purely two dimensional systems the sectional curvature appearing in equation 39 is given by $K = c\nabla^2 c - \vec{\nabla}c \cdot \vec{\nabla}c$. Common examples of depth dependant sound speed profiles used in modeling deep ocean acoustic wave guides appearing in standard texts may be analyzed using the Jacobi equation. In such situations all rays are torsion free and the set of rays in any vertical plane may be used to model the entire ray structure by rotation about the vertical axis.

For simplicity consider rays in the $x - z$ plane. For any ray an orthonormal basis may be constructed by inspection consisting of the unit vector along the ray $\vec{t} = \dot{x}\hat{i} + \dot{z}\hat{k}$ and two basis vectors, the first being obtained by rotation of the ray tangent by 90 degrees, $\hat{e}_1 = -\dot{z}\hat{i} + \dot{x}\hat{k}$, the other being $\hat{e}_2 = c\hat{j}$. In the conformal metric coordinate time is a proper affine parameter and "dot" refers to differentiation with respect to time. The basis vectors $(\hat{e}_1, \hat{e}_2)$ are orthonormal with respect to the conformal inner product $g_{ij} = \delta_{ij}/c^2$. Inspection of equation 81 shows that the sectional curvature in the $\langle \vec{t}, \hat{e}_1 \rangle$ plane is $K = cc'' - (c')^2$, where prime denote differentiation with respect to $z$, while the sectional curvature in the $\langle \vec{t}, \hat{e}_2 \rangle$ plane vanishes. From this it follows that rays fired from a point source at the same initial angle but in different planes will diverge linearly in time

The behavior of rays in a given vertical plane is governed by $K = cc'' - (c')^2$. The sound speed profile $c(z) = C\cosh(\sqrt{K_0} z / C)$ produces a space of constant positive curvature $K = K_0$. The deviation vector can be solved exactly for all rays in this system with the result $Y(t) = A\cos(\sqrt{K}t) + B\sin(\sqrt{K}t)$. One can conclude from this that periodic focusing will occur along each and every ray bundle with the same frequency regardless of the placement of the source. The time versus depth integral can be calculated in this case leading to the result that the travel time between turning points is independent of initial conditions, hence the point like focusing demonstrated in this case will always occur [Tolstoy].

For any depth dependant sound speed with the property that for some $z_0$ $c(z_0) = c_0$, $c'(z_0) = 0$ and $c''(z_0) = \alpha > 0$ one always has a single exact solution to the ray equation. Namely $z(t) = z_0$ and $x(t) = x_0 + c_0 t$, traveling along the waveguide axis. Along such a ray the basis may be chosen by inspection to be in the $z$ direction and the deviation equation solved immediately to give $Y(t) = A\cos(\sqrt{K}t) + B\sin(\sqrt{K}t)$ with $K = c_0\alpha$. From this result one can conclude that any ray sufficiently close to the wave guide axis of a focusing sound speed profile may be approximated by the coordinates $x_N(t) = x_0 + c_0 t$, $z_N(t) = A\cos(\sqrt{K}t) + B\sin(\sqrt{K}t)$. Clearly this rough estimate will have a better chance of approximating parabolic rays if the sound speed is a symmetric function about the horizontal axis defined by $z = z_0$. An approximation of this form is well suited for parabolic rays in a sound duct such as that produced by temperature and pressure gradients in the deep ocean.

The exact ray paths can be found for the sound speed $c(z) = a + bz$. The induced metric space is known in the literature as a Lobeschevsky space and is an example of a space of constant negative curvature. The deviation vector in this case may be found by inspection $Y(t) = A\cosh(\sqrt{K}t) + B\sinh(\sqrt{K}t)$, implying that all rays diverge from each other exponentially in time.

For stationary media the field amplitude in the high frequency limit is approximated by $A_t/A_0 = \sqrt{(\Delta\sigma_0/\Delta\sigma_t)(c_t/c_0)}$, where $A$ is the field amplitude, $\Delta\sigma$ the cress sectional area of a small tube surrounding the ray and $t$ is used as a subscript denoting evaluation of a quantity at a point along the ray at a given value of time. For a bundle of initially parallel rays the horizontal cross section remains constant and the ratio field amplitudes reduces to $A_t/A_0 = \sqrt{(Y_0/Y_t)(c_t/c_0)}$.

## V. DISCUSSION AND CONCLUSION

The general procedure for a paraxial ray tracing algorithm may be outlined as follows. Given the environmental parameters $\bar{\upsilon}(\bar{x},t)$ and $c(\bar{x},t)$, either as analytic functions or numerical data, determine a specific solution (or set of solutions) to equation 32 or 34 as the fiduciary ray (ray system) $\gamma_F(\lambda)$, $\gamma_N(\lambda)$. Solve equations 38, 43 or 45 to determine the parallel propagated basis at ever point along the fiduciary. This specific tangent field and space like basis defined on $\gamma_F(\lambda)$ along with the Riemann tensor components given in equations 40 through 42 are used to set up the equation for the deviation vector, equation 39 or 44. These equations may then be solved numerically, or in some cases analytically, to a desired accuracy for any initial conditions on the deviation vector, recalling that the initial conditions are chosen to model the initial shape of a small patch of the wave front. Given specific input values for the matrix components appearing in equation 39 one can check for the existence of convergence and divergence zones along the ray to determine if the ray is stable and if the procedure is expected to remain valid, within a desired accuracy.

The large number of solutions that may be found for layered media with a moving fluid are not only important for their pedagogical value in acoustic and differential geometry but may also provide decent approximations to real life situations in underwater or atmospheric acoustics. Especially in cases where there is fluid motion along with an approximately constant sound speed as equation 70 represents a formal solution in terms of a single integral.

**AKNOWLEDGEMENTS**

The author thanks the Office of Naval Research and the American Society for Engineering Education for hosting a summer faculty fellowship at the Naval Research Laboratory (NRL) in Washington DC for the summers of 2003 and 2004, during which time most of this work were completed. Portions of this written work were completed in part using resources provided by New York University through the Faculty Resource Network. The author would also like to acknowledge Dr. Daniel Wurmser, N.R.L., for sponsoring the fellowship and for collaboration and helpful comments during the start of the project. Finally special thanks to Dr. A. Lewis Licht for making the author aware of work by Matt Visser and William Unruh on acoustic analog models of black holes.

**REFERENCES**
[1]  V. Cerveny, Seismic Ray Theory, Cambridge University Press, 2001
[2]  C. W. Misner, K. S. Thorne and J. A. Wheeler, Gravitation, Freeman, New York, 1973
[3]  P. Schneider, J. Ehlers and E. E. Falco, Gravitational Lenses, Astronomy and Astrophysics Library, Springer-Verlag, New York, 1992
[4]  A. Abromowicz and W. Kluzniak, Epicyclic Orbital Oscillations in Newton's and Einstein's Dynamics, General Relativity and Gravitation, Vol. **35**, No. 1, pp. 69 – 77, 2003
[5]  W. G. Unruh, Experimental black hole evaporation?, Phys. Rev. Lett., **46**, 1351 – 1353 (1981)
[6]  M. Visser, Acoustic black holes: horizons, ergospheres, and Hawking radiation, arXiv:gr-qc/9712010, 1997
[7]  R. W. White, Acoustic Ray Tracing in Moving Inhomogeneous Fluids, J. Acoust. Soc. Am., Vol. **53**, No. 6, p. 1700 - 1704 (1973)
[8]  R. J. Thompson, Ray Theory for an Inhomogeneous Moving Medium, J. Acoust. Soc. Am., Vol. **51**, No. 5 (Part 2), p. 1675, 1972
[9]  O. V. Rudenko, A. K. Sukhorukova and A. P. Sukhorukov, Full Solutions to the Equations of Geometrical Acoustics in Stratified Moving Media, Acoustical Physics, Vol. **43**, No. 3, pp. 339 – 343 (1997)
[10]  L. Fishman, J. J. McCoy and S. C. Wales, Factorization and path integration of the Helmholtz equation: Numerical algorithms, J. Acoust. Soc. Am. **81**, 1355 – 1376 (1987)
[11]  D. Wurmser, G. J. Orris and R. Dashen, Application of the Foldy – Wouthuysen transformation to the reduced wave equation in range – dependant environments, J. Acoust. Soc. Am. 101 (3), 1997
[12]  S. W. Hawking and G. F. R  Ellis, The Large Scale Structure of Space – Time, Cambridge University Press, Cambridge, 1973
[13]  S . Kobayashi and K. Nomizu, Foundations of Differential Geometry Volume I and Volume II, John Wiley & Sons, INC., New York, 1963, (see Volume II specifically Chapter VIII for a discussion of Jacobi fields and conjugate point theorems)
[14]  B. O'Neill, Elementary Differential Geometry, Academic Press, INC., San Diego, 1966, (specifically chapter VII)
[15]  R. M. Wald, General Relativity, The University of Chicago Press, Chicago, 1984, (specifically chapters 1 – 3 for conventions, Appendix C and Appendix D are both referenced in section III.4.)
[16]  R. Guenther, Modern Optics, John Wiley & Sons, New York, 1990, (specifically chapter 5, see section entitled "Propagation in a Graded Index Optical Fiber" for a wave guide paraxial ray approximation)
[17]  R. Courant and D. Hilbert, Methods of Mathematical Physics Volume II, John Wiley & Sons, New York, 1962
[18]  Fritz John, Partial Differential Equations, Courant Institute of Mathematical Sciences Lecture Notes, New York University, New York, 1952
[19]  S. L.Bazanski, Kinematics of relative motion of test particles in general relativity (1), Ann. Inst. Henri Poincare, Section A, Vol. XXVII, No. 2, 1977, p. 155 – 144
27

**FIGURE CAPTIONS**

Figure1:
Relationship between the fluid velocity, ray tangent and wave front normal in Euclidian space.

Figure 2:
Sample of a complete space – time ray structure. (A) Family of geodesics labeled by *s*, where *s* parameterized flow lines are constructed from the deviation vector acting on a fiducial geodesic. (B) Each flow line represents a curve of constant affine parameter, λ. A tangent vector and the deviation at a single point are illustrated. (C) A curve of equal time, wave front, constructed from the intersection of a constant time plane with the geodesic flow. Equal parameter curves do not lie in the constant time surface in general.

Figure 3:
A fiducial geodesic and one of its neighbors are shown along with the fiducial tangent vector and the four dimensional deviation vector at two values of the affine parameter (labeled 0 and 1). The initial deviation vector is purely space like while the deviation vector at later times contains a time like component.

Figure 4:
Similar to figure 3, with the addition of the equal time deviation vector shown at both points along the ray (labeled 0 and 1). The four dimensional deviation, $Y$, and the three dimensional deviation, $\tilde{Y}$, are equal at the initial point by choice. As the system is tracked along $\gamma_F$ $Y$ picks up a time like component (as previously illustrated) while $\tilde{Y}$ contains only space components at all points.

Figure 5:
The shape of a small patch of the initial wave front and two neighboring wave front normal vectors are depicted for the three cases: (A) a concave, (B) convex and (C) a planar section of the wave front.

Figure 6:

A three dimensional ray path is shown along with the small ring formed by the locus of initial three dimensional deviation vectors. As the deviation vectors are propagated along the ray this ring deforms as a result of the local refractive index and fluid velocity. A pair of axis (labeled *a* and *b*) are included to demonstrate the change of orientation of the wave front.

Figure 7:
Convergence and divergence zones mapped in space for the symmetric fluid velocity described in this section with *n* = 1. Rays with $\alpha = 0$ travel in regions of zero sectional curvature (the result holding for any choice of $\upsilon$) (A). The Divergence zone, a region of negative curvature is defined by all rays fired against the wind, $\alpha < 0$, while an absolute convergence zone is defined by rays fined into the wind, $\alpha > 0$, with an initial angle $\cos\theta_0 > 2/3$. Additional divergence zones occur due to the dependence of *K* on the initial conditions.

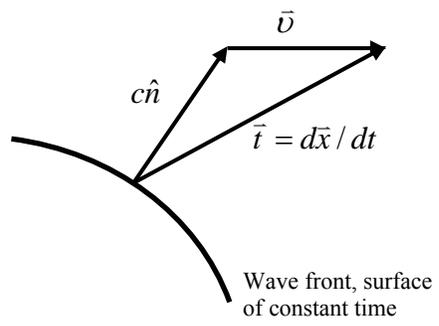

Figure 1

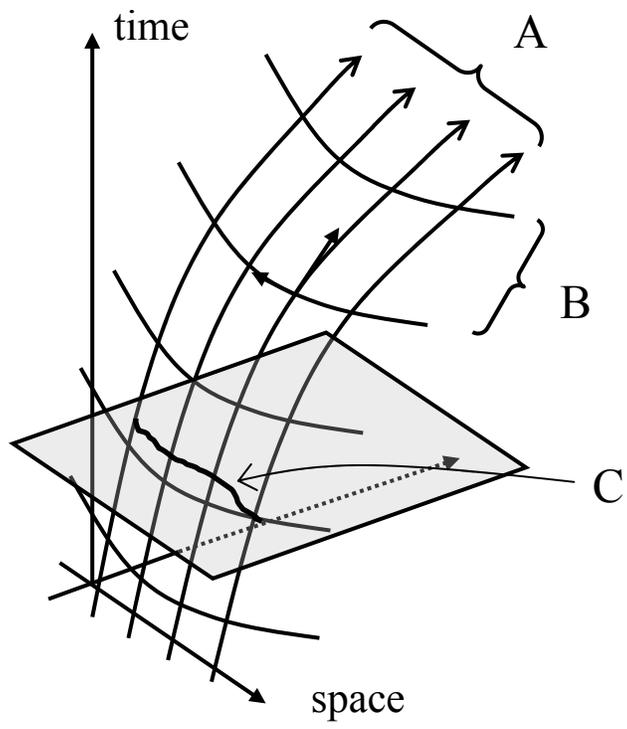

Figure 2

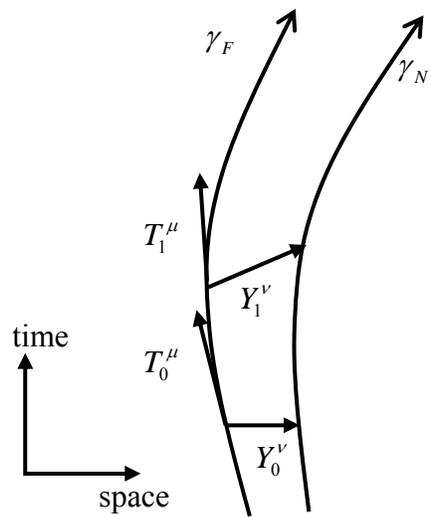

Figure 3

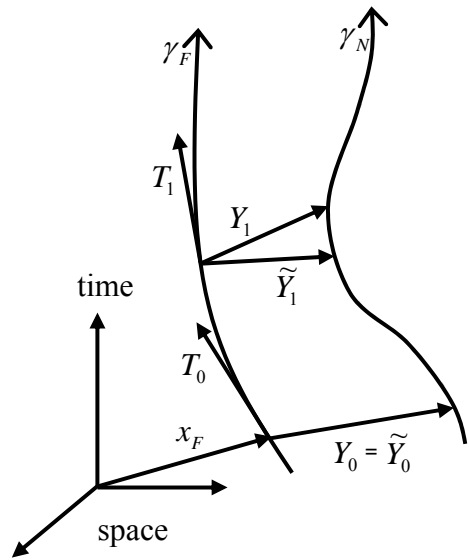

Figure 4

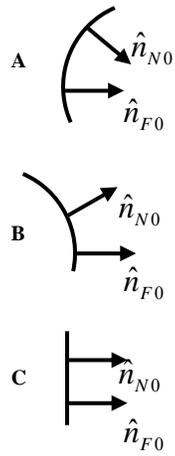

Figure 5

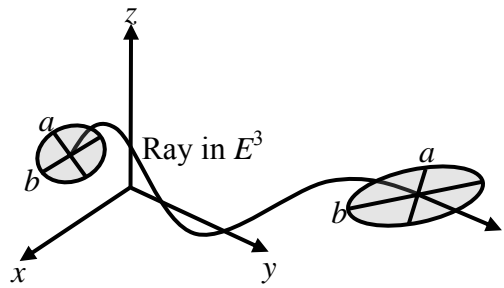

Figure 6

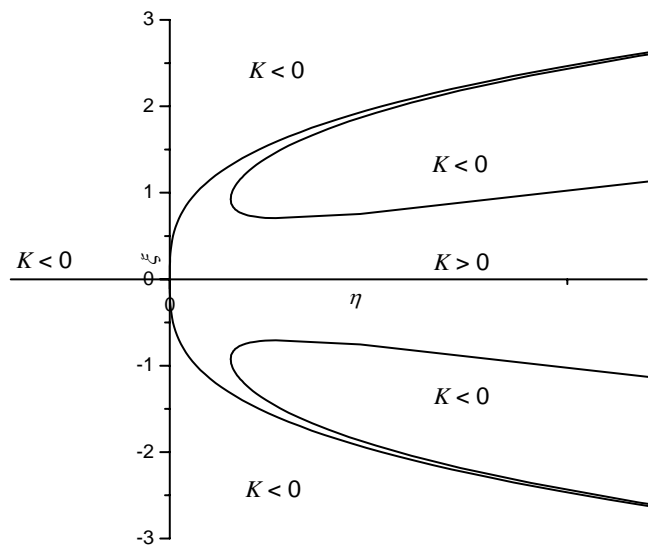

Figure 7